\documentstyle[12pt,a4wide,epsfig]{article}
\begin{document}

\renewcommand{\topfraction}{1}
\renewcommand{\bottomfraction}{1}
\renewcommand{\textfraction}{0}
\hfill \parbox[t]{3cm}{WUB 98-44 \\ \hfill HLRZ1998-85}
\begin{center}
{\Large\bf Flavor Singlet Axial Vector Coupling of the Proton \\
with Dynamical Wilson Fermions}
\vskip 1cm
SESAM Collaboration 
\vskip 0.5cm
S.~G\"usken$^a$, P.~Ueberholz$^a$, J.~Viehoff$^b$ \\

N.~Eicker$^b$, T.~Lippert$^a$, K.~Schilling$^{a,b}$, 
A.~Spitz$^b$, T.~Struckmann$^a$
\vskip 5mm
\normalsize\it{$^a$ Bergische Universit\"at Wuppertal, Fachbereich
  Physik,
  42097 Wuppertal, Germany} \\
\normalsize\it{$^{b}$ NIC Forschungszentrum J\"ulich, and DESY,
  Hamburg, 52425 J\"ulich, Germany} 




\abstract{\small We present the results of a full QCD lattice
calculation of the
flavor singlet axial vector coupling $G_A^1$ of the proton. The simulation
has been carried out on a $16^3\times 32$ lattice at $\beta=5.6$
with $n_f=2$ dynamical Wilson fermions.
It turns out that the statistical quality of the connected
contribution to $G_A^1$ is excellent, whereas the disconnected part
is accessible but suffers from large statistical fluctuations.  
Using a 1st order tadpole improved 
renormalization constant $Z_A$, we estimate $G_A^1 = 0.20(12)$.
}
\end{center}
\section{Introduction}
\normalsize
The flavor singlet axial vector coupling $G_A^1$ of the proton,
\begin{equation}
s_{\mu} G_A^1 = \langle P|
\bar{u}\gamma_{\mu}\gamma_5 u + \bar{d}\gamma_{\mu}\gamma_5 d +
\bar{s}\gamma_{\mu}\gamma_5 s| P\rangle
\label{eq_def_ga1}\quad,
\end{equation}
where $s_{\mu}$ denotes the components of the proton polarization
vector, has been the target of an intensive research activity both
in experimental and theoretical elementary particle physics over 
recent years. 
About one decade ago, the European Muon Collaboration (EMC) experiment
found an unexpectedly small value, $G_A^1=0.12(17)$, from
the measurement of the
first moment of the spin dependent proton structure function $g_p^1$
in deep inelastic polarized muon-proton scattering \cite{EMC_exp}.
Since $G_A^1$ can
be interpreted in the naive parton model as the fraction of the proton
spin carried by the quarks, the EMC result became known as the so called
`proton spin crisis'. 

A number of succeeding experiments have been
performed in the 
meantime \cite{SMC_exp,SLAC_exp1,SLAC_exp2,SMC_exp_new,SLAC_exp_new}. The
latest analysis\cite{SMC_exp_new}, including proton, neutron, and
deuteron data finds 
\begin{equation}
G_A^1 = 0.29(6) \quad,
\label{eq_ga1_expval}
\end{equation}
at a renormalization scale $\mu^2=5 \mbox{GeV}^2$. This is still
far away from the Ellis-Jaffe sum
rule\cite{ellis_jaffe_sr} expectation
$G_A^1 \simeq G_A^8 = 0.579(25)$\cite{ga8_expval}, which is found in
the Okubo-Zweig-Iizuka (OZI) limit of QCD.

As has been pointed out by Veneziano \cite{ga_anomaly1} a  
possible nontrivial topological structure of the QCD vacuum,
which induces a nonzero contribution 
to the divergence of the axial vector flavor
singlet current via the axial anomaly, might induce a sizeable
effect on the value of the matrix  
element in eq.(\ref{eq_def_ga1}). If one could prove that such vacuum
effects    
indeed reduce the value of $G_A^1$ from 0.579 to 0.29, then, on the
one hand, the `proton spin crisis' would be resolved since the naive
parton interpretation and the OZI estimate of $G_A^1$ would
be no longer valid. On the other hand,
such a result would establish the existence of nontrivial
topological vacuum structures, thus opening the door for a
quantitative study of the connection between vacuum topology and
instantons\cite{negele_review}.    
 
\begin{figure}
\begin{center}
\epsfxsize=12cm
\epsfbox{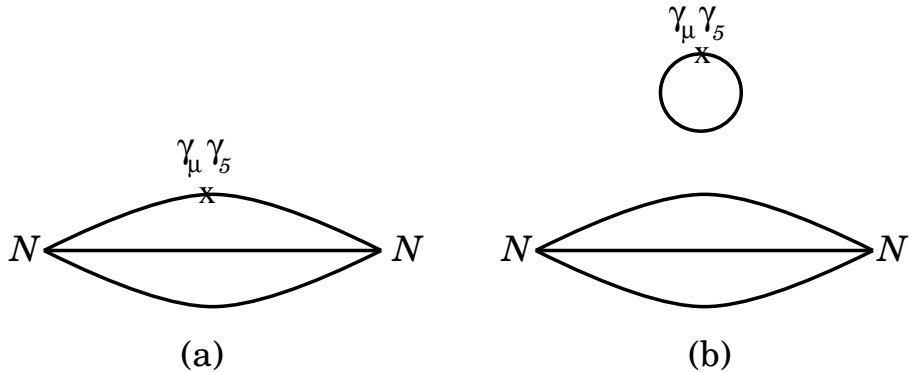}
\end{center}   
\caption{\label{fig_schema_con_dis}{\it Connected (a) and disconnected (b)
contributions to the axial vector density amplitude of a nucleon.
Please note that gluon lines connecting the quark lines are omitted. 
 }} 
\end{figure}

The lattice method provides an ideal tool to calculate 
the matrix element of eq.(\ref{eq_def_ga1}) nonperturbatively.
Within this method two complementary approaches have been developed 
over the recent years. The first is to  evaluate  the
connected and disconnected contributions to the flavor singlet
matrix element directly, which are schematically
depicted in fig.\ref{fig_schema_con_dis}. This ``direct method''
will be dealt with in the present paper. 
 
Alternatively, one can exploit the anomalous divergence equation
of the flavor singlet axial vector current.
Using this approach, the singlet coupling $G_A^1$ is extracted from the
correlation function of the proton propagator and the topological
charge density\cite{gupta_ga1,digiacomo_ga_quen,mtc_ga_full}. 
Although this ``topological method'' circumvents the computationally expensive
calculation of disconnected (fermionic) contributions, it
has some intrinsic difficulties. The connection between $G_A^1$ and
the correlation of proton and topological charge density is valid in the
limit of vanishing quark mass and zero momentum transfer only.
As both these limits cannot be taken by a direct calculation on a
finite lattice, two (wide range) extrapolations have to be performed.
In addition, the ``topological method'' is ill defined in the quenched
approximation of QCD\cite{gupta_ga1}. On the other hand,
the latter method provides the advantage of a nonperturbative
renormalization procedure for the topological charge 
density \cite{q_renorm_nonpert1,q_renorm_nonpert2}.   

The ``direct method'' has been tested on quenched configurations 
by Fukugita et al.\cite{japan_ga1} and Dong et al.\cite{liu_ga1} some
time ago. Although the former study has been done 
at a somewhat strong coupling ($\beta=5.7$) and the latter is based on
a very limited statistical sample (24 gauge configurations), the results
are encouraging, as they find estimates for $G_A^1$ compatible with    
eq.(\ref{eq_ga1_expval}).
 
It is of course by no means obvious, that these estimates hold also 
in full QCD. As we explained above, one expects vacuum
contributions to lower the parton estimate of $G_A^1$. The actual
numbers might well depend on the details of the vacuum
structure, such as the presence of sea quarks.   
Therefore, full QCD lattice simulations are necessary to calculate
reliable QCD results for $G_A^1$.
 
Exploratory full QCD studies using the ``topological method''
have been performed by the authors of ref.\cite{mtc_ga_full} and
ref.\cite{digiacomo_ga_full}, within the Kogut-Susskind
lattice discretization scheme and with $n_f=4$ degenerate quark flavors.
Both simulations suffer however from serious systematic uncertainties,
as they have been performed at finite quark mass and sizeable momentum
transfer ($q^2\simeq (600\mbox{MeV})^2$).

In this paper we present the results of a full QCD calculation
from the ``direct method''.

 We have analyzed 200 gauge configurations
at each of our four sea quark masses corresponding to
$m_{\pi}/m_{\rho}=0.833(3),0.809(15),0.758(11)$ and 0.686(11). 
The gauge configurations have been generated previously in the 
standard Wilson discretization scheme with $n_f=2$ mass degenerate
quark flavors,  at $\beta=5.6$ and with a lattice size of
$n_s^3\times n_t =16^3\times 32$ points.
Details of the simulation can be found in ref.\cite{sesam_light_spectrum}.

As a preparing remark we emphasize that the numerical calculation of
the disconnected parts of axial vector quantities is difficult,
especially in full QCD simulations.
It is known from lattice evaluations of the 
pion-nucleon-sigma term \cite{japan_nsigma,sesam_nsigma}
that the signals for the disconnected contributions of scalar
insertions suffer from large statistical fluctuations. Since the disconnected
parts of axial vector quantities can be seen as the difference of
two such scalar insertions, one can expect that the statistical noise
is even more dangerous in the axial vector case.
On top of this, a comparison
between quenched and full QCD lattice results for the light spectrum
revealed \cite{sesam_light_spectrum}, that the statistical
uncertainty in full QCD is larger  by
about a factor of 2 compared to a quenched calculation at equal
statistics, cutoff and lattice size.
Thus, one major issue of this
paper is to investigate whether and to what accuracy one can extract
disconnected signals for axial vector quantities with state of the art
methods and statistics in full QCD calculations.   

The paper is organized as follows. In the next section we will
briefly explain the methods used to calculate the connected and
disconnected contributions to the matrix element of
eq.(\ref{eq_def_ga1}) and present our raw data.
The chiral extrapolations and the renormalization are performed in
section 3. A discussion of
the physics results and concluding remarks are given in section 4.  

\section{Lattice Methods and Raw Data}

\subsection{Connected contributions}

To calculate the connected contributions to the axial vector density
matrix element of the proton, eq.(\ref{eq_def_ga1}), we have applied the
global summation method \cite{sesam_nsigma,maiani_ratio}. 
With this technique one calculates the ratio
\begin{eqnarray}
\lefteqn{R_{A_{\mu}}^{SUM}(t) =} \label{eq_sum_meth_def} \\
 && \frac{\sum_{\vec{x}}
\langle P^{\dagger}(\vec{0},0) \sum_{\vec{y},y_0}
\left[\bar{q}\gamma_{\mu}\gamma_5 q\right]
(\vec{y},y_0)
P(\vec{x},t) \rangle }
{\sum_{\vec{x}}\langle P^{\dagger}(\vec{0},0) P(\vec{x},t) \rangle}
-  \langle\sum_{\vec{y},y_0}
\left[\bar{q}\gamma_{\mu}\gamma_5 q\right](\vec{y},y_0)
 \rangle\;, \nonumber
\end{eqnarray} 
 where $P$ is an interpolating operator for the proton. From this
 ratio one extracts
the matrix element $\langle P|\bar{q}\gamma_{\mu} \gamma_5 q|P\rangle$,
which is given by the slope of the asymptotic form ,
\begin{equation}
R_{A_{\mu}}^{SUM}(t) \stackrel{t \rightarrow \infty}{\rightarrow}\;
A + \langle N|\bar{q}\gamma_{\mu}\gamma_5 q|N \rangle\,t \;. 
\label{eq_sum_meth_asympt}
\end{equation}   
Note that there are two different types of connected insertions, one
from the contraction of the ``loop'' quark $q$ with the $u$-type quark
of the proton  and one from the contraction with the $d$ type quark.  
This yields
\begin{equation}
R_{q,\mu}(t) = A_q + C_{q,\mu} t \quad,\quad 
C_{q,\mu} = \langle P(\kappa_{sea})|
\left[\bar{q}\gamma_{\mu}\gamma_5 q\right](\kappa_{sea}) 
|P(\kappa_{sea})\rangle_{con}\;,\quad q=u,d \;.
\label{eq_rq_cq_def}
\end{equation} 

The 3-point correlator in the numerator of eq.(\ref{eq_sum_meth_def})
has been evaluated using the standard insertion
technique\cite{maiani_ratio}. To improve on the ground state projection
of the proton we have applied 
`Wuppertal smearing' \cite{smearing_wtal} with $n=50,\alpha=4$ 
to all proton operators.

\begin{figure}[htb]
\begin{center}
\vskip -4cm
\epsfxsize=12.0cm
\centerline{\epsfbox{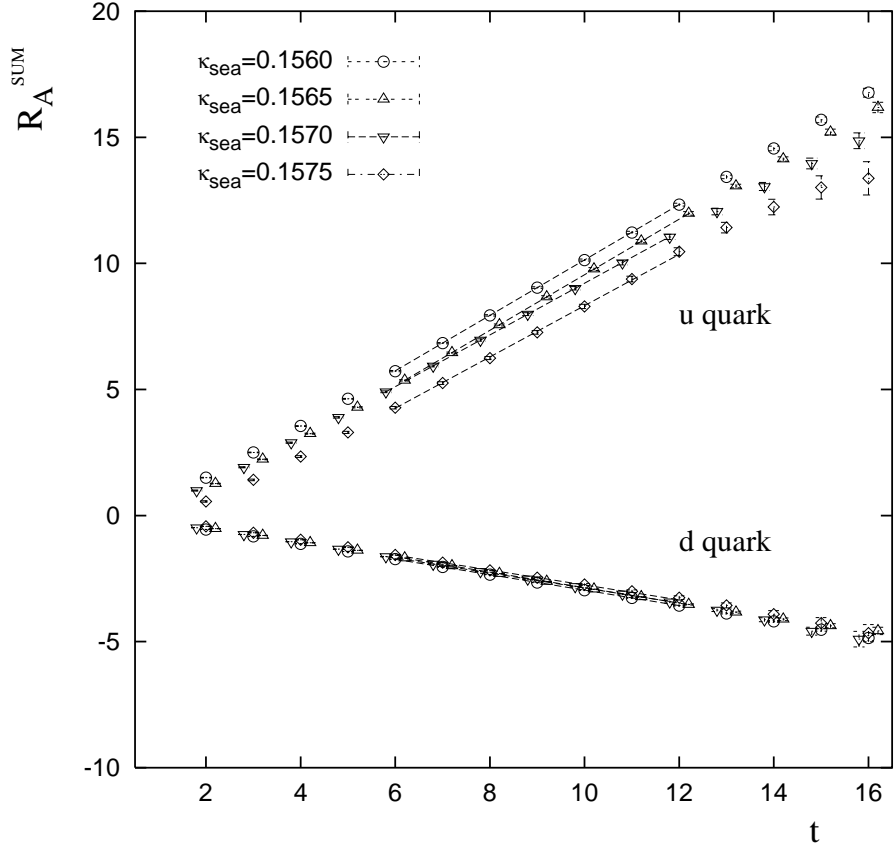}}
\caption{\label{fig_rsum_connected} {\it Summation method: The raw data $R_u$
  and $R_d$ for the  spin averaged connected contributions at our 
sea quark masses. The fits (range and value) are indicated by dashed lines.}}
\end{center}
\end{figure}

In fig.\ref{fig_rsum_connected} we display the corresponding signals,
spin averaged over $\gamma_{\mu}\gamma_5, {\mu}=1,2,3$. The sea quark
mass has been set equal to the (proton) valence quark mass. We
emphasize that
the expected linear rise in $R_A^{SUM}(t)$ is manifest
for all four quark masses.

The results of the fits to $R_A^{SUM}$ according to 
eq. (\ref{eq_rq_cq_def}) are collected in Table
\ref{tab_raw_con}. We find only a weak dependence on
the quark mass. 
\begin{table}[htb]
\begin{center}
\begin{tabular}{|c|c|c|c|}
\hline
$\kappa$ & $m_{\pi}/m_{\rho}$ & $C_u$ & $C_d$\\
\hline
 0.1560 & 0.833(5)  & 1.100(8)  & -0.307(3)  \\
 0.1565 & 0.809(15) & 1.102(11)  & -0.308(4)  \\
 0.1570 & 0.758(11) & 1.025(12)  & -0.297(8)  \\
 0.1575 & 0.686(11) & 1.018(24) & -0.295(10)  \\
\hline
\end{tabular}
\caption{\label{tab_raw_con}{\it Lattice results for the connected 
spin averaged amplitudes 
$C_u$ and $C_d$ at finite quark mass. 
}}
\end{center}
\end{table}
Note that the statistical errors to the connected amplitudes, which
have been determined using the jackknife method, are below 5$\%$. 
 
\subsection{Disconnected contributions}
 
The disconnected contributions to the axial vector density matrix
element  can be analyzed in principle with the global summation
method too. It has been pointed out however in ref.\cite{sesam_nsigma} that
the application of this technique induces large statistical fluctuations. 
We have  therefore chosen a slightly modified procedure, the plateau
accumulation method (PAM)\cite{sesam_nsigma}, which leads to a much
better signal to noise ratio.   
The PAM ratio is defined as
\begin{equation}
R_{A_{\mu}}^{PAM}(t,\Delta t_0,\Delta t)
 = \sum_{y_0=\Delta t_0}^{t-\Delta t} R_{A_{\mu}}^{PLA}(t,y_0)\;,
\label{eq_mplateau_meth_def}
\end{equation}  
where
\begin{eqnarray}
\lefteqn{R_{A_{\mu}}^{PLA}(t,y_0) =} \label{eq_plateau_meth_def} \\
 && \frac{\sum_{\vec{x}}
\langle P^{\dagger}(\vec{0},0) \sum_{\vec{y}}
\left[\bar{q} \gamma_{\mu}\gamma_5 q\right]
(\vec{y},y_0)
P(\vec{x},t) \rangle }
{\sum_{\vec{x}}\langle P^{\dagger}(\vec{0},0) P(\vec{x},t) \rangle}
- \langle\sum_{\vec{y}}\left[\bar{q}\gamma_{\mu}\gamma_5 q\right](\vec{y},y_0)
 \rangle \;. \nonumber
\end{eqnarray} 
$\Delta t$ and $\Delta t_0$ can be varied in the range
$1 \leq \Delta t,\Delta t_0 \leq t$ to study the influence of 
contributions from excited proton states. 

The asymptotic time dependence of $R_A^{PAM}$ is given by
\begin{equation}
R_{A_{\mu}}^{PAM}(t,\Delta t_0,\Delta t) =
B +  D_{q,\mu} (t-\Delta t - \Delta t_0) \;,
\label{eq_mplateau_meth_asympt}
\end{equation}
where $D_q$ denotes the disconnected contribution to the axial vector
amplitude
\begin{equation}
D_{q,\mu} = \langle N |\bar{q} \gamma_{\mu}\gamma_5 q|N \rangle_{disc} \;.
\label{eq_dq_def}
\end{equation}
The determination of the numerator of eq.(\ref{eq_plateau_meth_def})
requires the calculation of the correlation of the proton propagator
and the axial vector quark loop 
$L_{A_{\mu}}(y_0)=Tr(\bar{q}\gamma_{\mu}\gamma_5 q)$ at a given
timeslice $y_0$.
$L_A(y_0)$ has been estimated using the spin explicit stochastic
estimator technique \cite{sesam_nsigma} with complex $Z_2$ noise \cite{liu_z2}
and 100 stochastic estimates per spin component and configuration.
\begin{figure}[htb]
\begin{center}
\vskip -3.0cm
\noindent\parbox{15.0cm}{
\parbox{7.0cm}{\epsfxsize=7.0cm\epsfbox{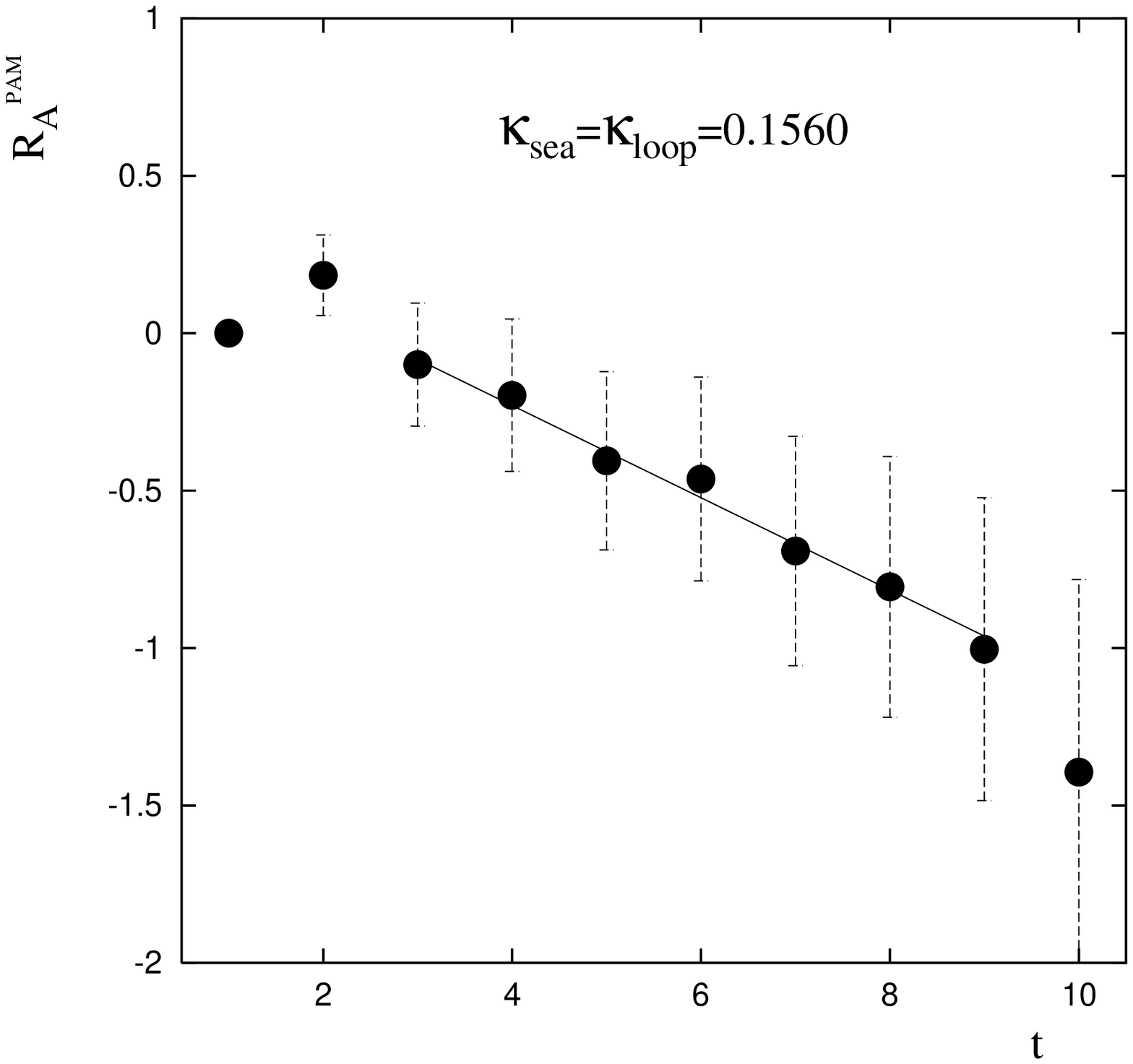}}
\parbox{7.0cm}{\epsfxsize=7.0cm\epsfbox{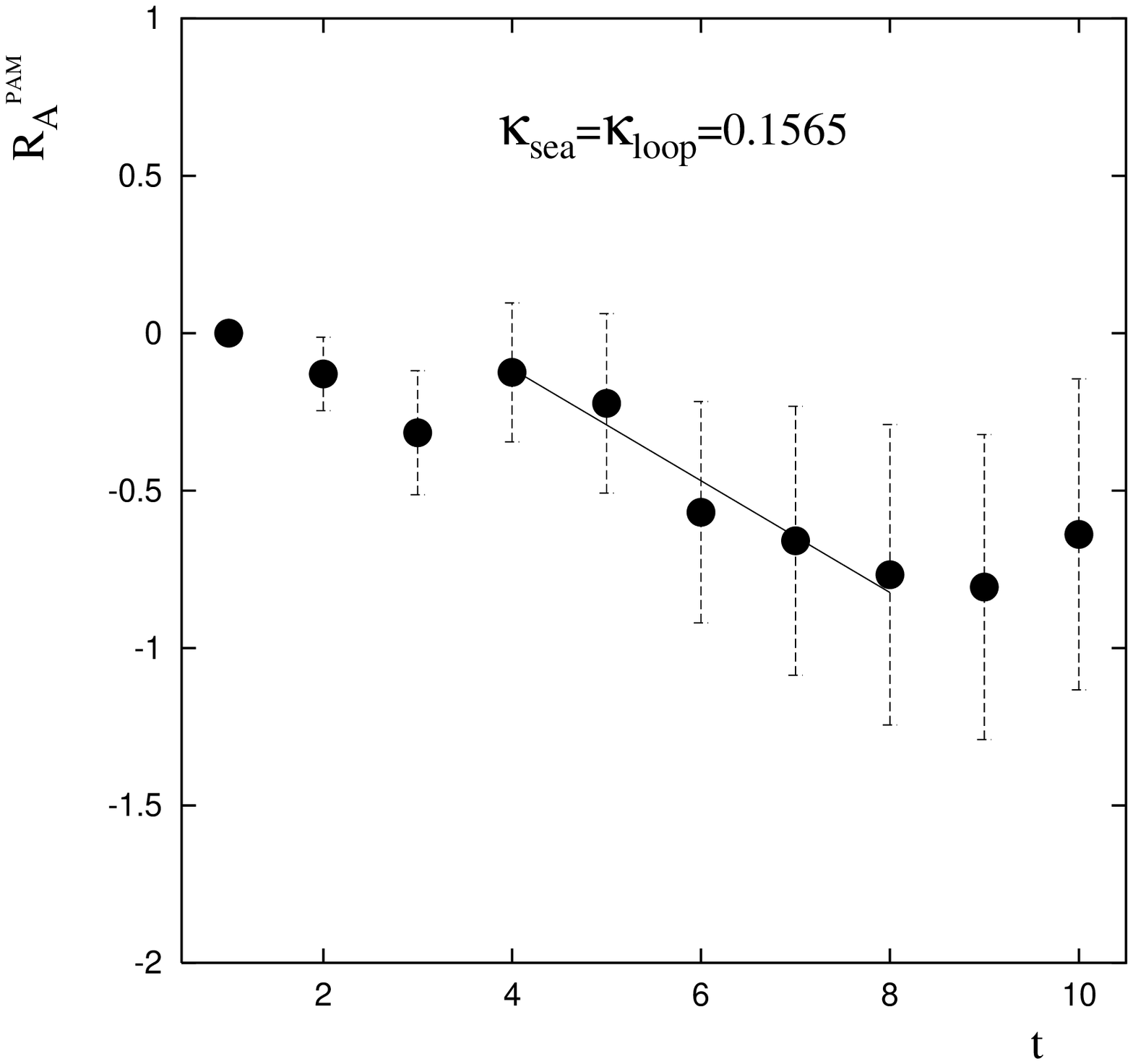}}
\\ \linebreak
\vskip -4.0cm
\parbox{7.0cm}{\epsfxsize=7.0cm\epsfbox{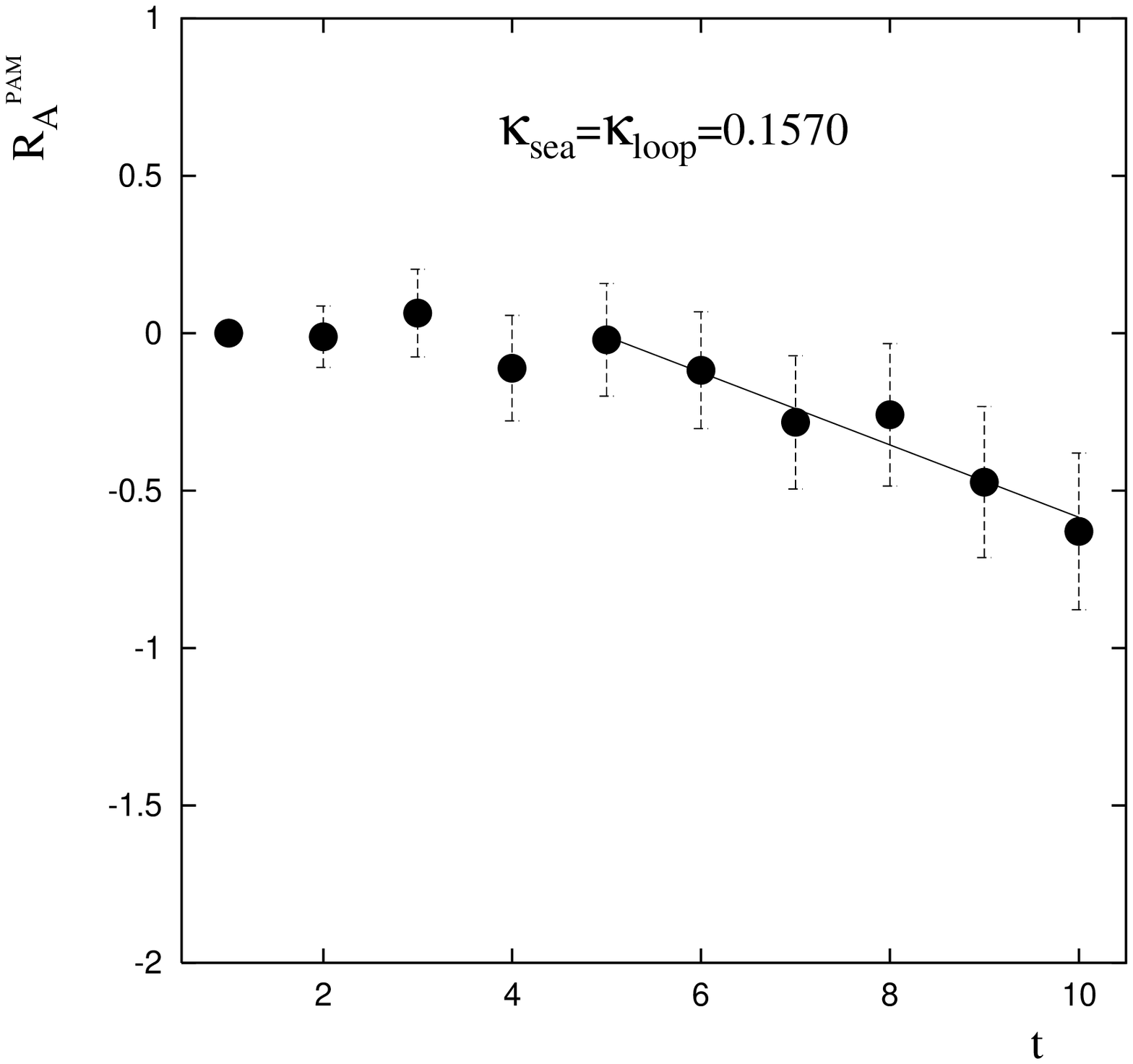}} 
\parbox{7.0cm}{\epsfxsize=7.0cm\epsfbox{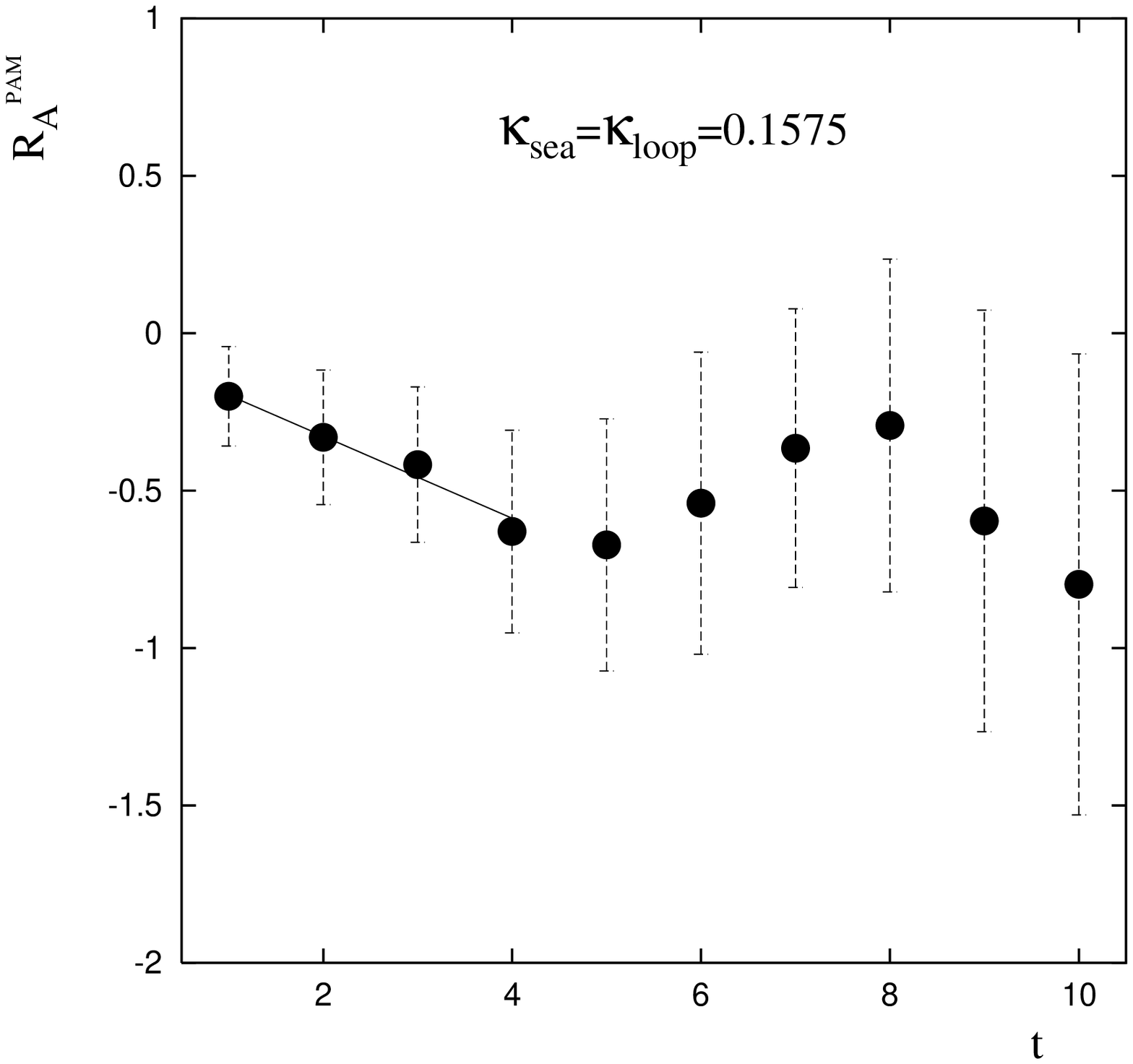}}\\
}
\caption{\label{fig_pam_disconnected} {\it The ratio $R_{A_3}^{PAM}(t)$ with
$\Delta {t_0}=\Delta t=1$ for the disconnected symmetric amplitudes 
$D_{q,3} =\langle P(\kappa_{sea})|
(\bar{q}\gamma_3\gamma_5 q)(\kappa_{sea})| P(\kappa_{sea})\rangle$ at our
sea quark masses. The fits (range and value) are indicated by solid lines.}}
\end{center}
\end{figure}

\begin{figure}[htb]
\begin{center}
\vskip -3.0cm
\noindent\parbox{15.0cm}{
\parbox{7.0cm}{\epsfxsize=7.0cm\epsfbox{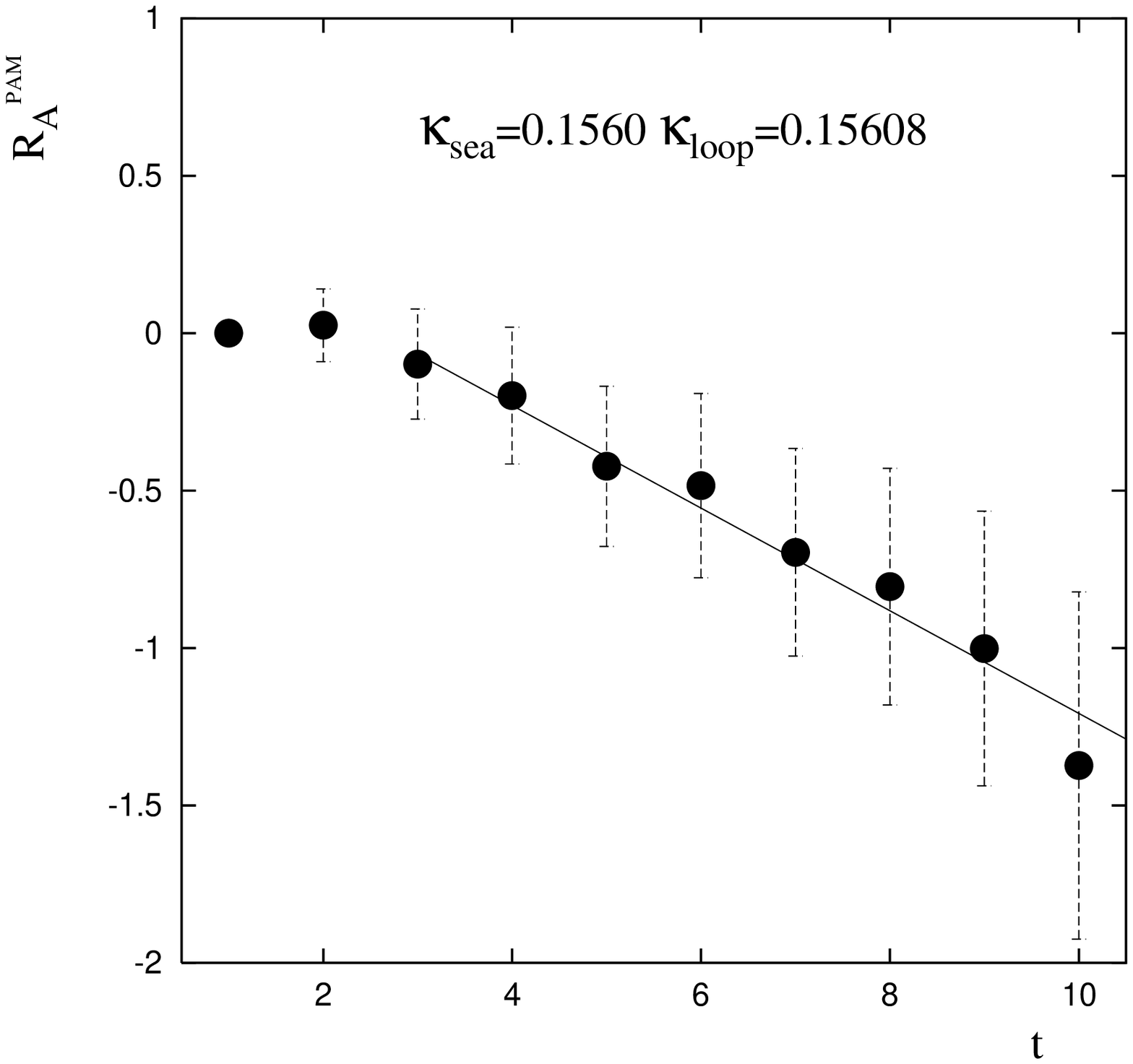}}
\parbox{7.0cm}{\epsfxsize=7.0cm\epsfbox{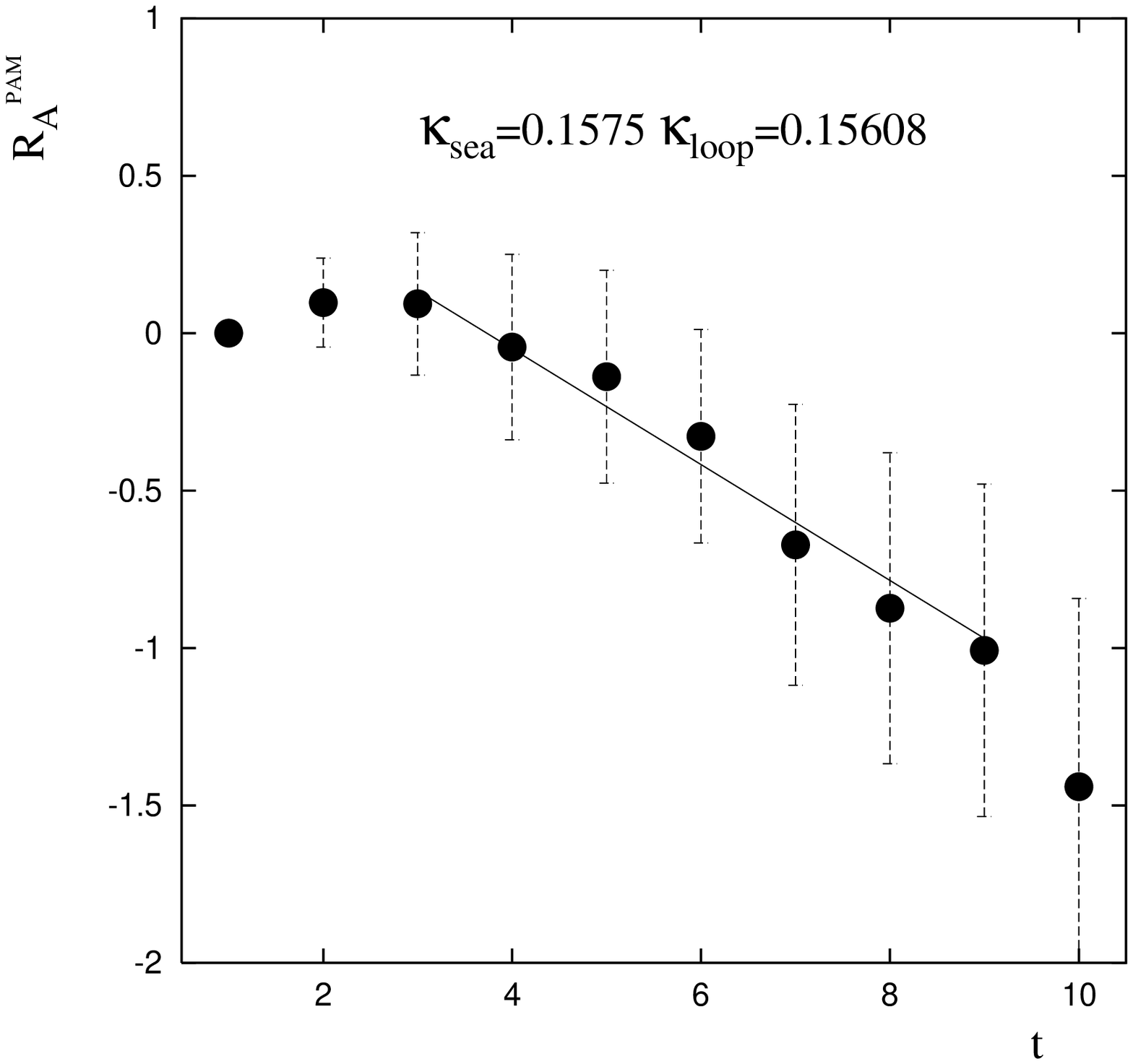}}
}
\caption{\label{fig_pam_disconnected_strange}{\it The ratio
$R_{A_3}^{PAM}(t)$  with $\Delta {t_0}=\Delta t=1$ for the disconnected
non symmetric amplitudes 
$D_{s,3} =\langle P(\kappa_{sea})|
(\bar{q}\gamma_3\gamma_5 q)(\kappa_{s})| P(\kappa_{sea})\rangle$ at
the heaviest and the lightest sea quark mass, and $\kappa_s=0.15608$.
The fits (range and value) are indicated by solid lines.}}
\end{center}
\end{figure}

We display in fig.\ref{fig_pam_disconnected} the signals\footnote{Note
that  we have used only the combination 
$\gamma_3\gamma_5$ in eq. (\ref{eq_plateau_meth_def}), since we found
the signals corresponding to the combinations $\gamma_1\gamma_5$ and
$\gamma_2\gamma_5$ to be dominated by statistical noise.}
$R_{A_3}^{PAM}(t)$, 
with\footnote{The signals $R_A^{PAM}(t)$ with $\Delta t_0 = \Delta t
>1$ are too noisy to extract a reliable signal.} $\Delta t_0 = \Delta t =1$,
for the symmetric case, where the (proton) valence quark mass and the
``loop'' quark mass have been set equal to the sea
quark mass.
Note that the statistical errors are quite large. 
In particular for our lightest quark mass,
corresponding to $\kappa_{sea}=0.1575$,
the signal is seen to be weak. We emphasize however,
that the slope of $R_A^{PAM}$ appears not to depend strongly on 
the quark mass.
Thus, excluding the
result for the lightest quark mass from the analysis would
not change the extrapolation to the chiral limit within errors.  
  
A similar behavior is found for the nonsymmetric case, where we have
set the `loop' quark mass equal to the strange quark
mass\cite{sesam_light_spectrum}, while keeping the valence quark mass 
at the sea quark mass. We show the corresponding signals  
in fig. \ref{fig_pam_disconnected_strange} for  the heaviest and 
the lightest quark mass. Within sizeable statistical errors one
can identify the expected linear decrease reasonably
well\footnote{The same situation is found at
$\kappa_{sea}=0.1565$. For $\kappa_{sea}=0.1570$ however, we observed 
a linear decrease for small $t$, though the asymptotic behavior is not
clear.}.

In Table \ref{tab_raw_dis} we have collected the raw results for 
$D_{q,3}$ and $D_{s,3}$ at the different quark masses. 
Apparently, the disconnected contribution to $G_A^1$ is definitely
less than zero, albeit  within  statistical errors of the order of
50$\%$. 
\begin{table}[htb]
\begin{center}
\begin{tabular}{|c|c|c|c|}
\hline
$\kappa$ & $m_{\pi}/m_{\rho}$ & $D_q$ & $D_s$\\
\hline
 0.1560 & 0.833(5)  & -0.146(60)  & -0.163(58)   \\
 0.1565 & 0.809(15) & -0.177(97)  & -0.125(74)  \\
 0.1570 & 0.758(11) & -0.119(62)  & -0.129(82)  \\
 0.1575 & 0.686(11) & -0.130(69)  & -0.184(76)  \\
\hline
\end{tabular}
\caption{\label{tab_raw_dis}{\it Lattice results for the 
disconnected amplitudes 
$D_{q,3} = \langle P(\kappa)|
[\bar{q}\gamma_{3}\gamma_5 q](\kappa)| P(\kappa)\rangle_{dis}$ and
$D_{s,3} = \langle P(\kappa)|    
[\bar{q}\gamma_{3}\gamma_5 q](\kappa_s)|
P(\kappa)\rangle_{dis}$. The value of the hopping parameter
$\kappa_s=0.15608$ corresponds to the strange  quark 
mass \cite{sesam_light_spectrum}.}
}
\end{center}
\end{table}

\section{Results}

\subsection{Chiral Extrapolations}

In order to arrive at our lattice estimate for $G_A^1$, we have to
extrapolate the connected ($C_u$, $C_d$) and the disconnected
amplitudes ($D_q$, $D_s$) to the light quark mass. Since the ``loop''
quark mass in $D_s$ is kept fixed at the strange quark mass, this 
quantity serves as a semiquenched estimate for the strange quark
contribution to $G_A^1$.
 
\begin{figure}[htb]
\begin{center}
\vskip -4cm
\epsfxsize=11.0cm
\centerline{\epsfbox{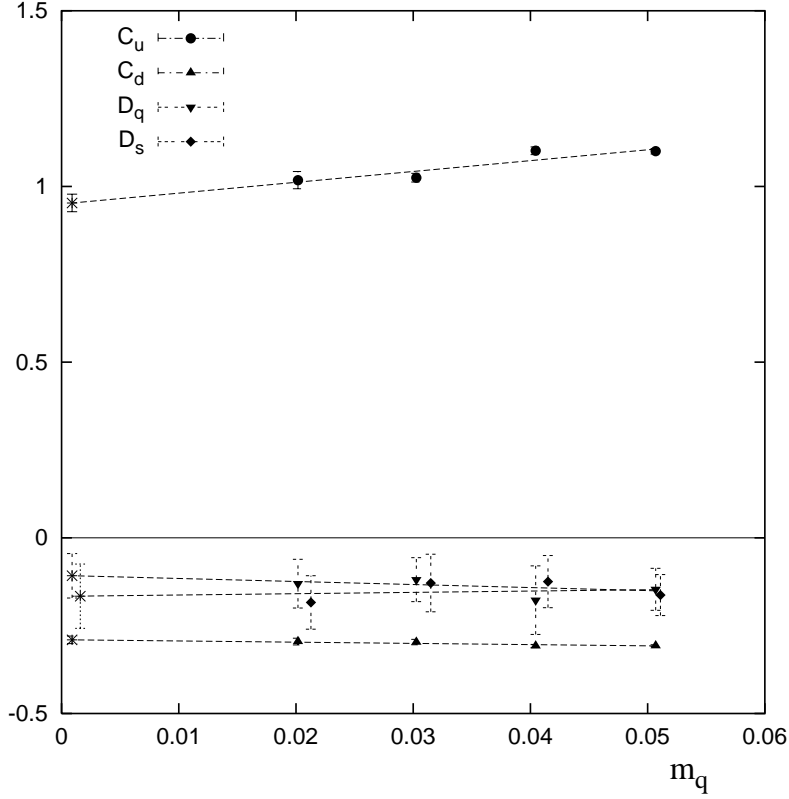}}
\caption{\label{fig_extrap_light} {\it Extrapolation of the
unrenormalized lattice amplitudes to the light quark mass. 
The fits are indicated by dashed lines, the results of the fits
by bursts. To avoid cluttering the data $D_s(m_q)$ have been
slightly shifted.}}
\end{center}
\end{figure}

We display the (unrenormalized) lattice data and the corresponding
fits as a function of the (sea) quark mass in
fig.\ref{fig_extrap_light}.
As mentioned above, the mass dependence of all the amplitudes is weak
and the linear ansatz for the extrapolations is in accord with our data.
Within errors, $D_q$ and $D_s$ do agree. We conclude that, to  the present
level of accuracy, flavor symmetry of the disconnected parts
appears to be  maintained in ($n_f=2$) full QCD. 
Note that the disconnected contribution to $G_A^1$, given by $2 D_q +
D_s$, is larger than the connected contribution $C_d$, suggesting
a sizeable modification to the OZI inspired estimate of the singlet coupling. 

At the light quark mass, being determined by the requirement that 
the ratio $m_{\pi}/m_{\rho}$ on the lattice equals
the experimental result \cite{sesam_light_spectrum}, we find
\begin{eqnarray}
C_u = 0.953(25) &\;,\;& C_d = -0.290(11) \;,\;\nonumber \\
                &     &  \label{eq_amp_at_mlight} \\
D_q =-0.108(63) &\;,\;& D_s = -0.166(91) \;. \nonumber
\end{eqnarray}
The numbers in brackets denote the statistical errors, obtained 
by a jackknife analysis of the extrapolation.
\begin{table}[htb]
\begin{center}
\begin{tabular}{|c|c|c|}
\hline
& & \\
$\Delta u = C_u + D_q$ & $\Delta_d = C_d + D_q$ & $ \Delta_s = D_s$ \\
& & \\
\hline
 0.85(10)  & -0.398(97)  & -0.166(91)   \\
\hline
& & \\
$G_A^1/Z_A^S = C_u + C_d + 2D_q + D_s$ & $G_A^3/Z_A^{NS} = C_u - C_d$
 & $G_A^8/Z_A^{NS} = C_u + C_d$ \\
& & \\
\hline
0.28(16)   & 1.243(28)   &  0.663(25)   \\
\hline 
\end{tabular}
\caption{\label{tab_raw_couplings_mlight}{\it Lattice results at the
light quark mass for the contribution of each quark flavor
and for the axial vector couplings of the
nucleon. ($Z_A^{NS})$,$Z_A^S$ denotes the (non)singlet axial vector
renormalization constant  
.}}
\end{center}
\end{table}

From the appropriate combinations of these amplitudes one can extract
the lattice values of the singlet and nonsinglet couplings, as well as  
the contributions of each single quark flavor to these quantities.
The results are collected in 
Table \ref{tab_raw_couplings_mlight}. Note that the nonsinglet
couplings have been calculated assuming flavor symmetry of the
disconnected parts.

\subsection{Renormalization and Physics Results}

In continuum QCD the flavor nonsinglet current $A_{\mu}^{NS}$ is 
conserved in the chiral limit. Thus it needs no renormalization.   
The flavor singlet axial vector current $A_{\mu}^{S}$ on the other hand 
is not conserved due to the presence of the axial anomaly. 
This leads in the continuum to a nontrivial
renormalization of $G_A^1$. In second order perturbation theory  
$A_{\mu}^{S}$ picks up a logarithmic divergence in the cutoff, which
induces a renormalization scale dependence on the axial flavor singlet
coupling, $G_A^1= G_A^1(\mu^2)$. The associated anomalous dimension
is known to 3 loops\cite{ga_renorm_cont}. Fortunately, it turns out
that the $\mu^2$ dependence of $G_A^1$ is very weak, e.g. increasing
$\mu^2$ from $10\mbox{GeV}^2$ to infinity decreases the value of
$G_A^1$ by about 10\% only, see  ref. \cite{SMC_exp_new}.

The situation is slightly more involved in the Wilson
discretization of lattice QCD \cite{karsten_smit}.
Already the nonsinglet current $A_{\mu}^{NS}$ is not conserved and 
has to be renormalized by a finite factor $Z_A^{NS}$. On top of this
the singlet current suffers from the axial anomaly. The resulting
scale dependence of $Z_A^S$ might be different from the
continuum form, as the axial anomaly mixes with the 
Wilson term.

In full QCD both, $Z_A^{NS}$ and $Z_A^{S}$, are known
only to first order (tadpole improved) lattice perturbation
theory\cite{sw_tadpole_imp,groot_renorm}, where they agree. For the nonsinglet
current, nonperturbative renormalization procedures have been developed and
used in the quenched
approximation\cite{luescher_nonpert,giusti_nonpert}.
Those methods, when 
applied to the case of full QCD, will yield a reliable value for
$Z_A^{NS}$.
 
The $\mu^2$ dependence of $Z_A^{S}$ however, which occurs beyond
first order, has not been calculated yet.  
One may hope of course, that it is as weak as in the continuum.

Thus, in view of the current state of analysis, we have used 
the first order tadpole improved estimate for $Z_A^{NS} = Z_A^{S}$,
\begin{equation}
Z_A^{NS} = \frac{1}{2\kappa}
\left( 1 - \frac{3\kappa}{4\kappa_c} \right)
\left( 1 - 0.31 \alpha_{\overline{MS}}(\frac{1}{a}) \right)\;,
\end{equation} 
with $\alpha_{\overline{MS}}(\frac{1}{a})=0.215$, $\kappa_c=0.158507$, and 
$\kappa=\kappa_l = 0.15846$, $\kappa=\kappa_s=0.15608$ for the light
and strange quark insertions respectively \cite{sesam_light_spectrum}.

Unfortunately this choice makes it difficult to determine the scale
$\mu^2$ at which we calculate $G_A^1$. The renormalized results
are compiled in Table \ref{tab_ren_couplings_mlight}.
\begin{table}[htb]
\begin{center}
\begin{tabular}{|c|c|c|c|c|c|}
\hline
$\Delta u$ & $\Delta_d$ & $ \Delta_s$ & $G_A^1$ & $G_A^3$  & $G_A^8$ \\
\hline
& & & & &\\
 0.62(7)  & -0.29(6)  & -0.12(7) & 0.20(12) & 0.907(20) & 0.484(18)   \\
& & & & &\\
\hline
\end{tabular}
\caption{\label{tab_ren_couplings_mlight}{\it Renormalized results at the
light quark mass for the contribution of each quark flavor
and for the axial vector couplings of the
nucleon. For the definitions of the quantities see
Table \ref{tab_raw_couplings_mlight}  
.}}
\end{center}
\end{table}
Comparing the estimates for $G_A^1$ and $G_A^8$ one finds that the
disconnected contributions lead to a violation of the Ellis-Jaffe
sum rule by more than 50$\%$, which is consistent with the experimental
finding. 
Note that there is a 30$\%$ discrepancy between our estimate of the
triplet coupling $G_A^3$ and the experimental value, 
$G_A^3 = 1.2670 \pm 0.0035$\cite{ga3_exp}. This points to the
presence of sizeable higher order or even nonperturbative
contributions to $Z_A^{NS}$. An uncertainty of similar size in $Z_A^1$
would however be well covered by the statistical errors of 50$\%$ in $G_A^1$.

\section{Discussion and Outlook}

We have calculated connected and disconnected contributions to the
flavor singlet axial vector coupling of the proton in a full QCD
$n_f=2$ lattice simulation with Wilson fermions. 
We find $G_A^1$,
within large uncertainties, to be consistent with the experimental
result and with previous quenched estimates.
This indicates that the substantial violation of the Ellis-Jaffe sum rule
might be caused mainly by the gluonic properties of the vacuum.  

The major sources of uncertainty are the large statistical
fluctuations of the disconnected parts and the use of the 1st order
estimate for $Z_A^S$. 

The statistical quality of disconnected axial vector insertions,
achieved with state of the art stochastic estimator
methods and with a sample size of 200 configurations, is encouraging
but not yet satisfactory. 
Ideas to improve on this situation are to use larger lattices, where self
averaging effects will help, to adapt the stochastic estimator
techniques to the axial vector case, e.g. by use of correlated noise
and, last not least,
by an increased statistics. Work along
these lines is in progress.

The nonperturbative determination of $Z_A^S$ presents a major problem.
One possible solution is to perform a nonperturbative renormalization of the
topological charge and then to use the `topological method' to
extract the flavor singlet axial vector renormalization constant.

Having overcome these problems one has to study cutoff and volume
dependence of $G_A^1$. Finally, as a future project, one should
perform full QCD lattice simulations with $n_f \geq 3$ non degenerate
dynamical fermions. This would allow for a realistic estimate of 
$D_s$ and of the amount of flavor symmetry breaking of the disconnected
contributions.
 
\paragraph{Acknowledgments}
This work has been supported by the DFG grants Schi
257/1-4, 257/3-2, 257/3-3 and by the DFG Graduiertenkolleg
``Feldtheoretische Methoden in der Statistischen und
Elementarteilchenphysik''. The connected contributions have been
computed on the CRAY T3E systems of ZAM at FJZ. The disconnected parts
were calculated on the APE100 computers at IfH Zeuthen and on the
Quadrics machine provided by the DFG to the Schwerpunkt "`Dynamische
Fermionen"', operated by the University of Bielefeld. We thank the
staffs of these institutions for their kind support.

\end{document}